# Fast Recursive Coding Based on Grouping of Symbols

Nikolay Ponomarenko, Vladimir Lukin, Karen Egiazarian, *Senior Member , IEEE,*
Jaakko Astola, *Fellow, IEEE,* and Boris Ryabko, *Member, IEEE*

*Abstract —* A novel fast recursive coding technique is proposed. It operates with only integer values not longer 8 bits and is multiplication free. Recursion the algorithm is based on indirectly provides rather effective coding of symbols for very large alphabets. The code length for the proposed technique can be up to 20-30% less than for arithmetic coding and, in the worst case it is only by 1-3% larger.

*Index Terms*—Arithmetic coding, Huffman codes, data compression, fast algorithms.

## I. INTRODUCTION

DATA compression is the area of intensive research during recent decades. Alongside with design of new coding methods using more sophisticated context modeling, an actual task is fast coding method design. Arithmetic coding (AC) proposed by Rissänen [1], in opposite to easier realizable Huffman coding (HC) [2], provides considerably less code redundancy. However, essential computations in AC, especially for adaptive modeling [3], restrict its use in applications requiring high coding speed, e.g. in video-data compression.

This stimulates designing different fast multiplication free AC algorithms [4,5] and fast algorithms for AC with adaptive modeling [6,7]. Another direction is the design of new coding methods faster than AC but having larger code redundancy [8].

Recently an approach to coding speeding-up dealing with each symbol division into two parts where only one is coded (i.e. using AC) and the other is simply numerated has appeared. Then one has to code symbols of considerably less size alphabets than for original alphabet, this leads to coding speeding-up.

Within this approach, two directions can be distinguished: the Moffat's K-flat codes [9] and Ryabko's techniques based on forming super-letters from symbols having almost equal occurrence probabilities [10,11]. If symbols can be divided into prefixes (beginnings) and suffixes (endings), then for Moffat's codes equal length prefixes (they are numerated) and different length suffixes (that are coded) are used. Ryabko's approach assumes such alphabet symbol grouping into super-letters where super-letters have different lengths (coded) and all symbol suffixes for given super-letter have equal lengths (numerated). This difference is illustrated by Fig. 1.

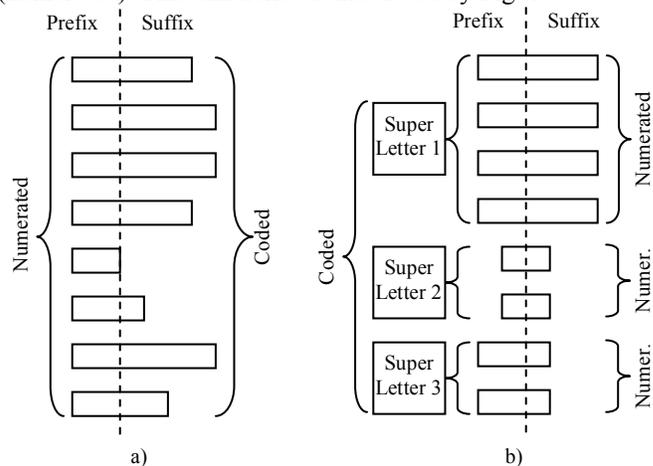

Fig 1. Coding simplifying by symbol division into coded and numerated parts
a) Moffat's approach  б) Ryabko's approach

Both approaches provide considerable speeding-up with not essential increasing of code-length, and have their own advantages. Moffat's approach allows performing fast search in compressed data, but for our case Ryabko's approach seems preferable. It can be successfully applied to block coding [12] and combined to both static and adaptive modeling [3] without increasing computations for used coding method.

Below, based on Ryabko's approach, we propose a novel effective recursive coding technique (further RCGS - Recursive Coding based on Grouping of Symbols) which implies only symbol numeration. This technique operates with only data not larger 8 bits. For this technique there is no necessity in using AC or another "external" coding at final stage as for methods in [9,11].

## II. RECURSIVE CODING BASED ON GROUPING OF SYMBOLS

### A. Basic idea

Let's divide original text into two streams: super-letter

Manuscript received ????.

N. Ponomarenko and V. Lukin are with the National Aerospace University, 61070, Kharkov, Ukraine (e-mail: lukin@xai.kharkov.ua).

K. Egiazarian and J. Astola are with Institute of Signal Processing, Tampere University of Technology, Tampere, FIN-33101, Finland (e-mails: karen@cs.tut.fi, jta@cs.tut.fi)

B. Ryabko is with the Siberian State University of Telecommunication and Computer Science, Novosibirsk, 630102, Russia (e-mail: ryabko@niec.nsk.su).

stream and suffix index stream (one way of doing this see in *B*). The latter stream is saved to output data as it is. Super-letter stream is quite "heterogeneous" and it should be somehow compressed. In fact, super-letter stream is also the text that differs from original text by less alphabet size. The idea of the proposed coding techniques is the following. If the condition

$$N \geq N_s^2, \qquad (1)$$

where *N* is the original alphabet size, $N_s$ denotes super-letter alphabet size, is valid, then after pair-wise uniting all data of super-letter stream into data with twice larger size, the obtained alphabet size is not larger than *N*. The text length due to grouping in pairs of neighbor data decreases twice. If this procedure is repeated recursively for new text with satisfying (1) at each step, then after each step the length of data remained non-coded decreases by two times. After few steps ($\leq \log_2 L$, where *L* is the original text size) all text will be coded. In Fig. 2 this idea is represented as block-diagram.

Note that for RCGS the symbol probability table should not be stored in output data stream for static modeling as for AC. For decoding one has to know only the super-letters number and their content. RCGS with adaptive modeling is also possible, but as the basis for RCGS we consider just static modeling that ensures maximal speed of data encoding/decoding.

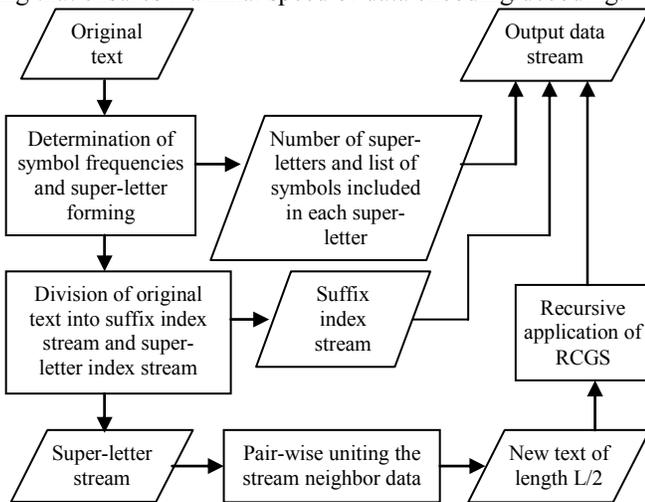

Fig.2. RCGS block diagram

### B. Practical realization

While designing practical RCGS realization one has to select *N* and the method of symbol grouping to super-letters. The alphabet with $N = 2^{16}$ suits only for adaptive modeling since the list of symbols grouped into super-letters is too large for storing in coded data although the algorithm of grouping [11] provides $N_s = 253$ (under condition (1)) with appropriate additional code redundancy 2.6%.

For the alphabet with $N = 2^8$ the realization of RCGS with static modeling is possible. However, the grouping algorithm [11] produces required $N_s = 16$ with possible additional code redundancy with approximately 23%. Obviously such additional redundancy is too large.

But the algorithm [11] assumes that we know only alphabet size. Below we describe an algorithm for symbol grouping into super-letters for known symbol probabilities as it is in static modeling.

While grouping alphabet symbols into super-letters, the code length of these symbols increases. This increase in normalized form can be expressed as

$$\Delta = -p_s(-\log_2 p_s + \log_2 M) / \sum_{i=1}^{M}(p_i \log_2 p_i), \qquad (2)$$

where *M* is the number of symbols grouped into super-letter, $p_s$ is their aggregate probability $\sum_{i=1}^{M} p_i$.

Let the symbols are grouped into super-letter if $\Delta$ does not exceed some acceptable threshold $T_\Delta$. Then the grouping algorithm can be the following:
1. Set $T_\Delta$, e.g. $T_\Delta = 0.01$;
2. Sort alphabet symbols in ascending order according to their probabilities;
3. Consider in descending order all possible *M* equal to powers of 2 (for alphabet with $N = 2^8$ these *M*=256, 128, 64, 32, 16, 8, 4, 2, and 1). For each *M* check $\Delta \leq T_\Delta$ (take the first *M* symbols from the sorted sequence). First time the condition becomes valid, group symbols into super-letter and remove them from further forming of super-letters;
4. If there are symbols not yet grouped to super-letters, repeat step 3.

Time spent on super-letter forming for this algorithm does not depend on *L* and in most practical situation has small contribution into total coding time. We have not met practical situations when for $T_\Delta = 0.02$ $N_s \leq 16$. If this happens, one has to increase $T_\Delta$ a little and repeat super-letter forming.

### C. Coding/decoding procedure aspects

In text coding after forming super-letter set, original text is divided into super-letter stream and suffix index stream. Due to use of static modeling, for each symbol the super-letter index, suffix index and suffix length in given super-letter are known. The bit number for super-letter index for $N = 2^8$ never exceeds 4, suffix index bit number can not be larger 8. Then it seems reasonable to create small tables with each symbol correspondence to super-letter index, suffix index and length. For symbol coding, the operations consist in taking super-letter and suffix indices from the table and their passing to the corresponding streams. This is very fast.

Then the super-letters are grouped in pairs in super-letter stream. This operation is performed as simple shift of the first grouped super-letter $S1$ by 4 bits and logic «or» with the second super-letter $S2$: $Sr = (S1 \; shl \; 4) \vee S2$ with getting $Sr$ as their grouping result.

The algorithm is recursive, but due to decreasing the coded symbol number twice at each step the total number of coded symbols does not exceed $2L$. Per each coded symbol of original text one needs, on the average, not more than two extractions from table and not more than one shift and one "or" operations. Two additions are needed for symbols probability calculation per each text letter.

Decoding is performed in inverse manner: grouped super-letter pairs are divided as $S1 = Sr\ shr\ 4$, $S2 = Sr \wedge 240$; then for each super-letter the table of correspondence of suffix indices to original text symbols is formed using saved information on super-letter content.

## III. NUMERICAL SIMULATIONS

RCGS basic advantages with respect to AC are higher coding speed with negligibly larger code redundancy. Besides, due to recursion of coding algorithm at each step of which symbol probabilities are taken into account, RCGS is able to effectively code symbols for very large alphabets.

These were the reasons to use as test data: the Calgary Text Compression Corpus files; sequences of integer valued variables with non-uniform distribution; and a text with very large alphabet (it was quantized DCT coefficients for 8x8 blocks of the images Lenna and Barbara). This corresponds to alphabet with $N = 2^{512}$ (each symbol length is 64 bytes).

For comparisons we used RCGS ($T_\Delta = 0.01$), AC and HC. For correct comparison of coding speed data, static modeling for AC and HC was used. For estimation of additional redundancy of RCGS, the entropy of original text was calculated. The used program of AC was well speed optimized and written in Assembler. RCGS software is written in Delphi with middle optimization.

Table I gives sample results for some files and the averaged results for all files. Let us draw attention to the following. The AC code length exceeds text entropy by approximately 3%. This deals with errors of fast integer valued realization of AC and, in less degree, with static modeling use (output data include probability table). RCGS provides, on the average, even less code length than text entropy (by 1.3 %) due to algorithm recursion that partly takes into account probabilities of symbol pairs, four symbol groups, etc. RCGS has, at least, twice smaller encoding/decoding time than AC. And there are still resources for RCGS software optimization.

TABLE I
Results for Calgary Text Compression Corpus files

| File | Per symbol Entropy | Per symbol code length | | | Encoding time, ms | | Decoding time, ms | |
|---|---|---|---|---|---|---|---|---|
| | | AC | HC | RCGS | AC | RCGS | AC | RCGS |
| bib | 5.201 | 5.312 | 5.429 | 5.184 | 23.96 | 11.83 | 24.99 | 11.72 |
| geo | 5.646 | 5.688 | 6.617 | 4.713 | 24.13 | 10.67 | 24.93 | 10.57 |
| news | 5.190 | 5.266 | 5.398 | 5.187 | 86.29 | 38.33 | 87.26 | 37.96 |
| paper3 | 4.665 | 4.833 | 4.844 | 4.725 | 11.93 | 5.21 | 12.23 | 5.16 |
| pic | 1.210 | 1.311 | 4.178 | 0.968 | 60.96 | 37.27 | 62.28 | 36.90 |
| Average | 4.891 | 5.042 | 5.407 | 4.826 | 36.80 | 17.89 | 38.05 | 17.72 |

The case of random uncorrelated (quantized with quantization step QS) data coding (Table II) is the most unfavorable for RCGS. But even in this case RGCS provides code-length only by 1.5-2% larger than entropy. The data in Table III outline RCGS ability to effectively code very large alphabet symbols. The code length for RCGS is 5-30% less than entropy.

TABLE II
Results for Gaussian Noise (quantized with QS=1, 256 Kb array) coding

| $\sigma^2$ | Per symbol Entropy | Per symbol code length | | | Encoding time, ms | | Decoding time, ms | |
|---|---|---|---|---|---|---|---|---|
| | | AC | HC | RCGS | AC | RCGS | AC | RCGS |
| 0.5 | 1.658 | 1.802 | 2.064 | 1.680 | 36.13 | 23.57 | 37.72 | 23.34 |
| 25 | 4.370 | 4.475 | 4.420 | 4.443 | 59.56 | 27.20 | 62.99 | 26.94 |
| 400 | 6.369 | 6.433 | 6.414 | 6.445 | 74.30 | 27.56 | 74.59 | 27.59 |

TABLE III
Results for quantized DCT coefficients coding (256 Kb array)

| Picture | QS | Per symbol Entropy | Per symbol code length | | | Encoding time, ms | | Decoding time, ms | |
|---|---|---|---|---|---|---|---|---|---|
| | | | AC | HC | RCGS | AC | RCGS | AC | RCGS |
| Lenna | 3 | 3.067 | 3.143 | 4.940 | 2.852 | 47.74 | 26.90 | 50.36 | 25.09 |
| | 30 | 0.711 | 0.837 | 4.101 | 0.494 | 37.13 | 17.78 | 37.69 | 17.63 |
| Barbara | 3 | 3.497 | 3.560 | 5.019 | 3.260 | 48.58 | 26.45 | 49.80 | 26.19 |
| | 30 | 1.042 | 1.166 | 4.140 | 0.803 | 35.71 | 19.32 | 37.41 | 19.13 |

## IV. CONCLUSION

This paper introduces a novel recursive coding technique applicable as fast, simple and effective alternative to AC. RCGS is applicable to 8 and less bit data, it is multiplication free. RCGS outperforms AC in coding/decoding speed by approximately twice and it often provides code length smaller than AC (up to 30%). In the worst cases the code lengths for RCGS are only 1-3% larger than entropy.

REFERENCES

[1] J. Rissänen, "Generalized kraft inequality and arithmetic coding," IBM J. Res. Develop., vol. 20, pp. 198-203, May 1976.
[2] D. A. Huffman, "A method for the construction of minimum-redundancy codes," Proc. Inst. Radio Eng., vol. 40, no. 9, pp. 1098-1101, Sept. 1952.
[3] T. Bell, I. H. Witten, and J. G. Cleary. Modeling for text compression. ACM Computing Surveys, 21(4), pp.557-591, Dec. 1989.
[4] J. Rissänen and K.M. Mohiuddin, "A multiplication-free multialphabet arithmetic code", IEEE Transactions on Communications, vol. 37, issue: 2 , pp. 93 – 98, Feb. 1989.
[5] D. Chevion, E.D. Karnin and E. Walach, "High efficiency, multiplication free approximation of arithmetic coding", Data Compression Conference, pp. 43-52, April 8-11, 1991.
[6] D. W. Jones, Application of splay trees to data compression, Communications of the ACM, 31:8, pp. 996-1007, 1988.
[7] B. Ya. Ryabko, A fast sequential code. Soviet Math. Dokl., 39:3(1989), pp. 533-537.
[8] U. Graf, "Dense coding - a fast alternative to arithmetic coding", Proceedings of Compression and Complexity of Sequences, pp. 295-304, June 11-13, 1997.
[9] M. Liddell and A. Moffat, "Hybrid Prefix Code for Practical Use", Procedeengs of Data Compression Conference, pp. 392-401, 2003.
[10] B. Ryabko and J. Rissänen, "Fast adaptive arithmetic code for large alphabet sources with asymmetrical distributions", IEEE Communications Letters, vol. 7, issue 1, pp. 33–35, Jan, 2003.
[11] B. Ryabko, J. Astola and K. Egiazarian, "Fast codes for large alphabets", Communications in information and systems, vol. 3, no. 2, pp. 65-78, October, 2003.
[12] B. Ryabko, G. Marchokov, K. Egiazarian, J. Astola, "The fast algorithm for the block codes and its application to image compression", Proceedings of ICIP, vol. 2, pp.205-207, Sept. 14-17, 2003.